\documentclass[11pt,twoside]{article}

\usepackage{asp2006}
\usepackage{epsf}
\usepackage{lscape}

\begin{document}
\title{The influence of the environment on bar formation}   
\author{J. M\'endez-Abreu$^{1}$;  J. A. L. Aguerri$^{1}$, S. Zarattini$^{2}$, R. S\'anchez-Janssen$^{1}$ \& E. M. Corsini$^{2}$}   
\affil{$^1$ Instituto   de  Astrof\'isica  de  Canarias,  $^2$
Dipartimento di Astronomia, Universita di Padova
}    

\begin{abstract} 
Galaxy mergers and interactions  are mechanisms which could drive the
formation of  bars.  Therefore, we  could expect that the  fraction of
barred galaxies  increases with the  local density.  Here we  show the
first results of an extensive  search for barred galaxies in different
environments.    We  conclude   that  the   bar  fraction   on  bright
(L$>$L$^{*}$)  field,  Virgo,  and  Coma   cluster  galaxies  is
compatible.   These  results  point  towards  an  scenario  where  the
formation and/or  evolution of bars  depend mostly on  internal galaxy
processes rather than external ones.
\end{abstract}


\section{Bar formation: nature vs. nurture}   
The observational proofs about the influence of the environment on bar
formation and/or  evolution are still rare.   In \citet{aguerri09}, we
investigate  a volume-limited  sample  of 2106  disc  galaxies in  the
nearby universe to derive the bar  fraction as a function of the local
galaxy density.  The local  density was
calculated for  every sample galaxy using the  fifth nearest neighbor
method,  obtaining that  80\% of  our  galaxies were  located in  very
low-density  environments  ($\Sigma_5   <$  1  Mpc$^{-2}$),  and  20\%
(corresponding to  more than 400  galaxies) covers mostly  the typical
values  measured for  loose ($\Sigma_5  >$ 1  Mpc$^{-2}$)  and compact
galaxy groups  ($\Sigma_5 \sim$  10 Mpc$^{-2}$). We  did not  find any
difference  between the local  galaxy density  of barred  and unbarred
galaxies in our  range of densities.  

To  extend this  conclusion to  higher density  environments,  we have
investigated the fraction of barred galaxies present in the two nearby
\emph{benchmark}  clusters: Virgo  (Zarattini et  al., in  prep.)  and
Coma (M\'endez-Abreu et al., in  prep.).  In both clusters, we found a
bar fraction in their bright galaxy population consistent with that in
the  field. Similar  results have  been recently  found in  cluster at
higher   redshift  by  Marinova   et  al.    (2009)  and   Barazza  et
al.  (2009). However,  the errors  in the  clusters bar  fractions are
large  and  the  results  might   be  affected  by  the  small  number
statistics. To deal with this  problem, we are undergoing an ambitious
project  to  study the  bar  fraction in  a  large  sample of  cluster
galaxies drawn from the WINGS project.





\begin{thebibliography}{}
\bibitem[Aguerri et al.(2009)]{aguerri09} Aguerri, J.~A.~L., M{\'e}ndez-Abreu, J., \& Corsini, E.~M.\ 2009, \aap, 495, 491
\bibitem[Marinova et al.(2009)]{marinova09} Marinova, I., et al.\ 
2009, \apj, 698, 1639 
\bibitem[Barazza et al.(2009)]{barazza09} Barazza, F.~D., et al.\ 2009, \aap, 497, 713 


\end{thebibliography}
\end{document}